\documentclass[twocolumn,preprintnumbers,amsmath,amssymb]{revtex4}
\usepackage{epsfig}
\usepackage{dcolumn}
\usepackage{bm}
\usepackage{color}

\begin{document}
\renewcommand{\thefootnote}{\fnsymbol{footnote}} 

\title{{On the} feasibility of  nanocrystal  imaging using intense and ultrashort {1.5 \AA} X-ray pulses\\}


\author{C. Caleman$^{1,2}$, G. Huldt$^3$, F. R. N. C. Maia$^3$, C. Ortiz$^4$, F. G. Parak$^1$, J. Hajdu$^3$, D. van der Spoel$^3$,
H.  N. Chapman$^{2,5}$
 and N. Timneanu$^3$}
\email[]{E-mail: nicusor@xray.bmc.uu.se}

\affiliation{$^1$Physik Department E17, Technische Universit{\"a}t M{\"u}nchen, James-Franck-Stra\ss e, DE-85748 Garching, Germany\\
$^2$Center for Free-Electron Laser Science, DESY, Notkestra\ss e 85, DE-22607 Hamburg, Germany\\
$^3$Department of Cell and Molecular Biology, Biomedical Centre,
Box 596, Uppsala University, SE-75\,124 Uppsala, Sweden\\
$^4$Institut f{\"u}r Theoretische Physik, Goethe-Universit{\"a}t Frankfurt, Max-von-Laue-Stra\ss e 1, DE-60438 Frankfurt am Main, Germany\\
$^5$ University of Hamburg, Luruper Chaussee 149, DE-22761 Hamburg, Germany}

\maketitle

\noindent 
{\bf Structural studies of biological macromolecules are severely limited by radiation damage.
Traditional crystallography curbs the effects of damage by spreading damage over many copies of
the molecule of interest. 
X-ray lasers, such as the recently built LINAC Coherent Light Source (LCLS)~\cite{DiMauro2007a}, 
offer an additional opportunity for limiting damage by out-running damage processes with ultrashort and very intense X-ray pulses.
Such pulses may allow the imaging of single molecules, clusters or nanoparticles, but coherent
flash imaging will also open up new avenues for structural studies on nano- and micro-crystalline substances.
This paper addresses the theoretical potentials and limitations of nanocrystallography with extremely intense coherent X-ray pulses.
We use urea nanocrystals as a model for generic biological substances and simulate primary and secondary ionization dynamics in the crystalline sample. Our results establish conditions for ultrafast nanocrystallography diffraction experiments as a function of fluence and pulse duration.}

\begin{figure} [t]
  \begin{center}
 \includegraphics[width=.9\columnwidth]{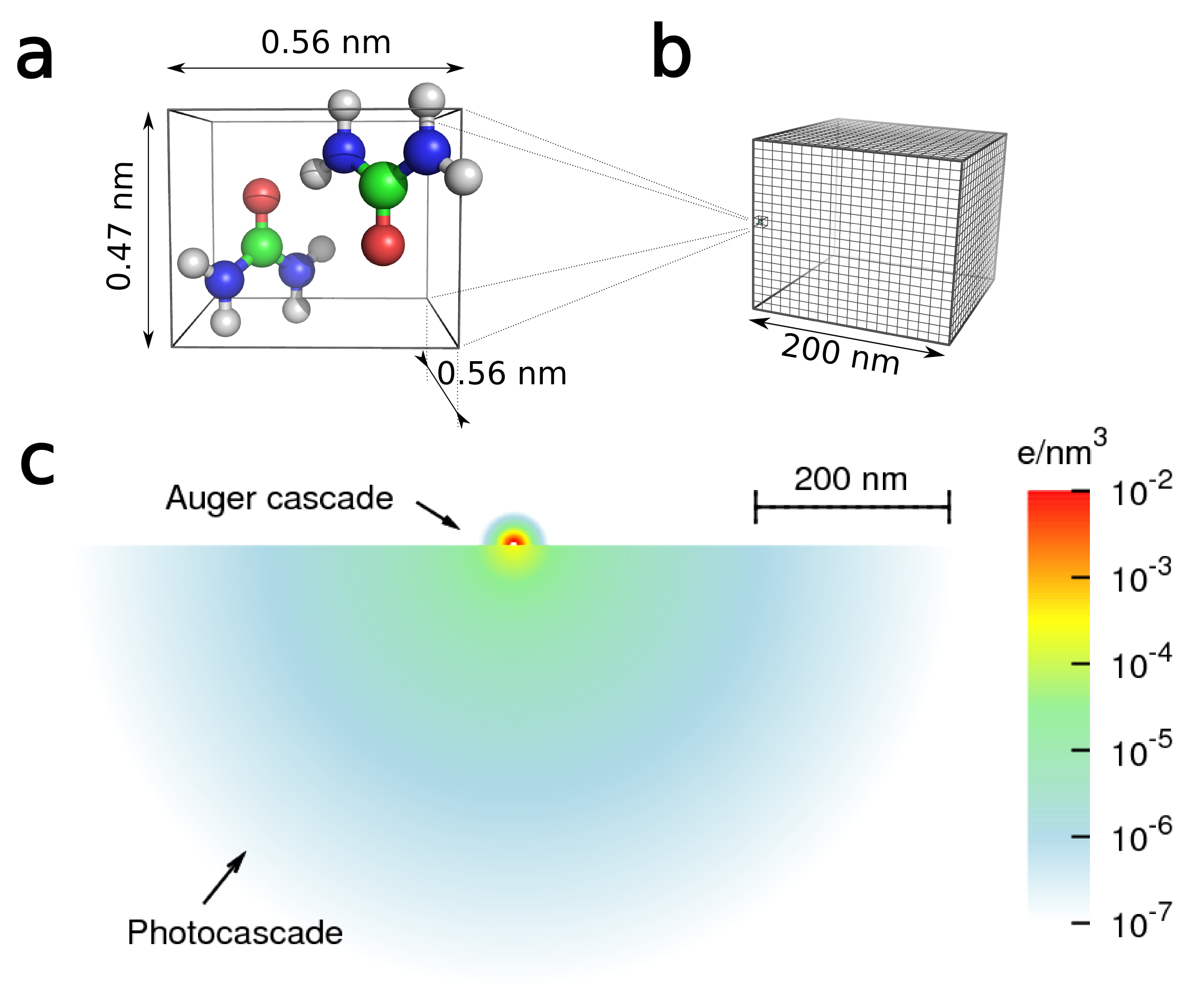}
     \caption{{\bf Comparison of crystal size and the {modeled} size of secondary electron cascades.} (a)~The unit cell of a urea crystal contains light elements abundant in proteins: carbon (depicted as green), nitrogen (blue), oxygen (red) and hydrogen (white). (b)~A urea nanocrystal of 200 nm would contain about 50 million unit cells. A protein nanocrystal of similar size would contain about 20,000 unit cells (using lysozyme as an example). (c)~The overall dimensions of simulated electron clouds produced during the thermalization of a single 0.4~keV Auger electron ejected from a nitrogen atom (top)
   and a single 8~keV photoelectron (bottom) inside a large urea crystal after 50 femtoseconds. Similar cascade sizes are produced in protein crystals, during an X-ray diffraction experiment. The total number of ionizations was 18 in the Auger cascade, and 390 in the photoelectron cascade at 50 fs after the emission of primary electrons. At this point, the radius of gyration of the photoelectron cloud reached 300~nm, and that of the Auger electron cloud 8~nm. The photoelectron cascade is bigger than a typical nanocrystal  under consideration here.}\label{fig:cloud_drawing}
  \end{center}
\end{figure}

Any sample exposed to an intense X-ray pulse
will be ionized and extensive ionization destroys the sample. 
The time scale on which this process occurs is critical for obtaining an interpretable diffraction pattern that conveys an atomic structure of the
sample. In principle, the X-ray pulse must be short enough for the
entire pulse to pass through the sample before
major disarrangement of atomic and electronic configurations takes
place. The
ionizations due to direct photoabsorption
 and subsequent secondary processes affect the ability to
get useful structural information from the
diffraction pattern in three ways:
(i)~Ionization 
decreases the elastic X-ray scattering power of atoms { and induces considerable changes in 
diffracted intensities due to ionization stochasticity}.
(ii)~Removal of electrons from atoms leaves behind positively charged
ions that repel each other due to Coulomb forces, leading to
the destruction of the structure.
(iii)~Free electrons 
either leave the sample, if their energy is high enough, or remain in the sample
as a background electron gas, in which case they will be a source of
noise in the diffraction image.

\begin{figure*}[t]
  \begin{center}
    \includegraphics[width=2.1\columnwidth]{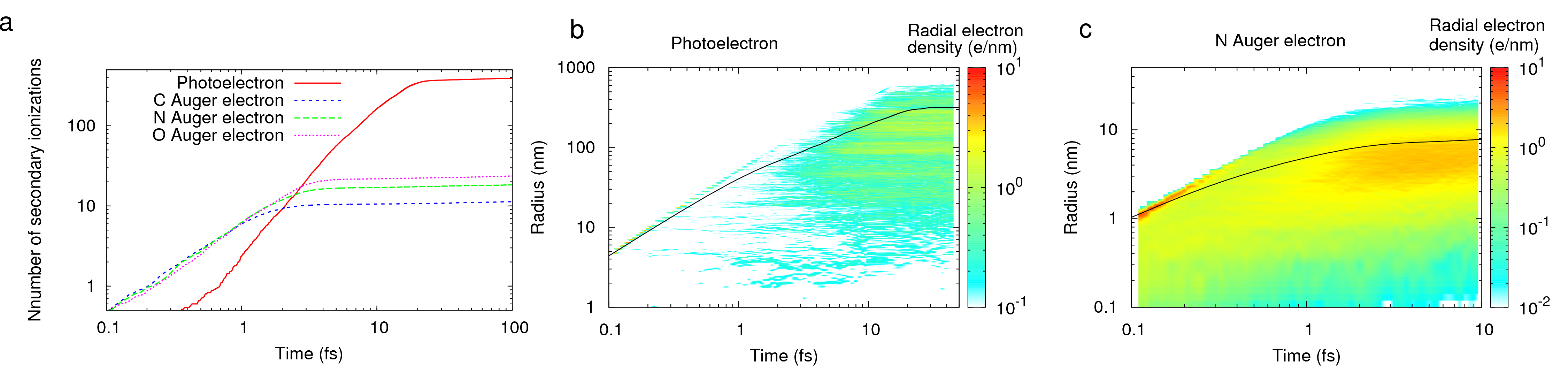}
   \caption{{\bf Evolution of secondary ionization cascades in a urea crystal over time.} (a)~Number of secondary ionizations produced by a photoelectron of 8 keV and by Auger electrons (impact energies: 250~eV for carbon, 400~eV for nitrogen, 500~eV for oxygen). 
   (a)~Spatial evolution of the electron cloud from a single photoelectron of 8 keV energy in an infinitely large urea crystal depicted through the radial electron density as a function of time and radial distance from the point of incidence. The cascade includes the primary electron and its secondary electrons. At any given time, the 4$\pi$ integration over the radius in the 3D volume gives the total number of {electrons}, assuming spherical symmetry when the cascades are added stochastically. (c)~Secondary electron cloud from a single Auger electron (nitrogen). The cascade includes the primary electron and its secondary electrons. The thermalization of electrons from oxygen and carbon has similar features.  Black lines in (b) and (c) show the radii of gyration (Equation~\ref{eq:rg}) of the electron clouds. In the first femtosecond the electron clouds are highly anisotropic. After 20 fs no more secondary ionizations will occur in the photoelectron cascade (5 fs for Auger cascades). 
  Figures show averaging over 1000 simulations on an infinitely large urea crystal. 
}\label{fig:ionAUGvsPHOTO}
  \end{center}
\end{figure*}

Thermalization of trapped electrons leads to additional ionizations through cascade processes. 
The probability of trapping, and the size of the resulting secondary electron cascades, depends on sample size and X-ray energy  (among others). 
Photoelectrons released by X-rays of 1.5~{\AA}~wavelength are fast 
(53~nm/fs) and can escape from small samples 
such as "nanosized" crystals~\cite{Nave2005b,Cowan2008a} 
early in an exposure (Figure~\ref{fig:cloud_drawing}). 
 In contrast, Auger electrons from carbon atoms are slow (9.5~nm/fs) and 
cause
 secondary ionization even in a single protein molecule~\cite{Bergh2004a,Caleman2009a}.
For small samples (diameter$<$10~nm), 
the explosion is dominated 
by the repulsion of positive ions left behind by electrons leaving the sample~\cite{Bergh2004a,Jurek2004b}. In big samples (diameter$>${500}~nm), most electrons will be trapped simply because they lose energy before reaching the surface. Trapped electrons increase the {temperature} of the sample through {collisional} processes, while slowing the explosion by partially screening the positive charges and creating a net neutral core. Predictions point to a transition from Coulomb explosion to a hydrodynamic expansion {with increasing sample size}. A positively charged surface layer is formed, destroying the sample from outside towards the {center}. The expansion 
is driven by thermal processes as the electron pressure grows~\cite{Riege2004a}. Hence, 
{crystalline diffraction and}
useful structural information might still be {obtained from the center of the crystal.} 

\begin{figure*}[t]
  \begin{center}
   \includegraphics[width=2.1\columnwidth]{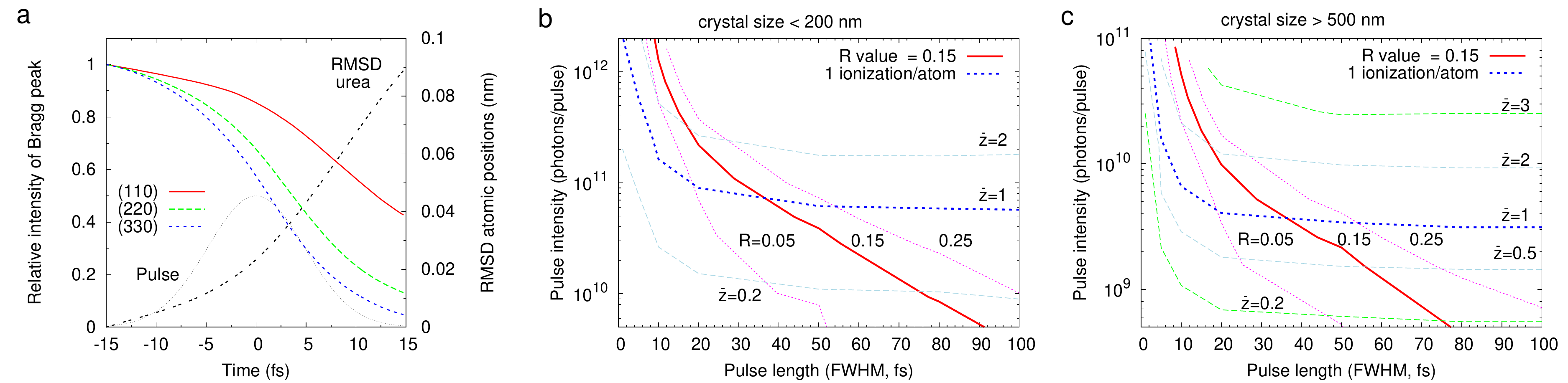}
    \caption{{\bf Degradation of the detected signal as a result of radiation damage.}
    (a) Decay of Bragg peaks during exposure to an X-ray pulse in a urea crystal. Pulse length: 10~fs (FWHM) centered at t=0, at 1.5~{\AA} wavelength. The pulse intensity (1.5$\times$10$^{11}$ photons in a focal spot of 1 $\mu$m diameter FWHM) is such that atoms are ionized once in average when 99.5\% of the intensity of the pulse passes through the sample. Peak intensities at different resolutions are represented by the (hkl) reflections and are normalized to one, based on intensities from undamaged crystals. The (330) reflection corresponds to 1.2~{\AA} resolution and has a pulse-integrated degradation of 50\% due to ionization and atomic displacement. The dashed black line shows the average root mean square deviations (RMSD) in atomic positions during illumination. (b,c) Contour plots for the average ionization per atom ($\bar z$) and the {R-value} as a function of the X-ray pulse length and intensity (photons/focal spot of 1 $\mu$m diameter FWHM). The {R-value} (Equation~\ref{eq:rfactor}) is calculated from all the Bragg peaks up to a resolution of 1.5~{\AA}. The red thick line corresponds to an {R-value} of {0.15}, lower values are considered acceptable for a good reconstructable signal. The blue thick dashed line represents the damage of 1 ionization per atom. The plot in (b) shows behavior of nanocrystals smaller than 200 nm, from which the photoelectrons escape early in an exposure, while (c) shows the behavior of crystals larger than 500 nm, when photoelectrons are completely trapped during exposure.}\label{fig:bragg}
  \end{center}
\end{figure*}

Descriptions of electron impact ionization and secondary ionization cascades 
exist for different materials~\cite{Caleman2009a,Ziaja2005a,Ortiz2007a}.
Dynamics of photoelectrons in protein crystals have been investigated earlier~\cite{Nave2005b,Cowan2008a}, 
without consideration to Auger emission or secondary ionization cascades. These predictions suggest
that radiation damage can be limited by reducing the crystal size.
The present work steps beyond these studies and gives an integrated description of photo-emission, 
Auger emission and cascade processes during exposure of a biological nanocrystal to an ultrashort and intense X-ray pulse, { to determine the feasibility of nanocrystal imaging and 
improvement in resolution achievable with shorter pulses. Our findings are summarized in Figures~\ref{fig:cloud_drawing}-\ref{fig:intensity} and the methodology is described in Methods and Supplementary}.

For light elements, a single photo-ionization releases electrons at two distinctly different energies (Figure~\ref{fig:cloud_drawing}). The photoelectron energy corresponds to the difference between the photon energy 
and the K-shell binding energy, while Auger electrons carry kinetic energy dependent on atom {type} (250 eV for carbon).  
The {average time for the first collisional ionization} scales with the primary electron energy (Figure~\ref{fig:ionAUGvsPHOTO}a). 
The electron cloud initiated by a photoelectron
thermalizes slower than electrons in Auger cascades, 
since energetic electrons travel further between scattering events in the
crystal due to their lower interaction cross section (Figures~\ref{fig:ionAUGvsPHOTO}b,c). 
At the same time, the cloud generated by a photoelectron is around four orders of magnitude 
larger in volume than the Auger electron induced cloud.
After thermalization, the electron clouds keep expanding through diffusion, following a random walk pattern.
Figures~\ref{fig:ionAUGvsPHOTO}b,c show that
 the radius of gyration (Equation~\ref{eq:rg}) at these impact energies
 describes well the spatial extent of the electron clouds.

{In a sample that is small compared to the size of the X-ray beam or the photon absorption length, photoionization
events will occur with equal probability throughout the entire sample.}
At 8.3~keV photon energy, a single photoelectron will liberate about 400 electrons before reaching thermal equilibrium (Figure~\ref{fig:ionAUGvsPHOTO}a). 
The electron gas will have high temperature due to the high photon
energy, 
and the electrons will be distributed 
approximately uniform
throughout the sample.
The free electrons will scatter predominantly in the forward direction
and contribute incoherently to the background in the diffraction pattern.




At 1.5~\AA~wavelength, the ratio between elastically 
scattered photons and photoionization is 1:32 for oxygen, 1:26 for nitrogen and 1:20 for carbon. 
Incoming photons will primarily ionize sample and only a few will contribute to coherent scattering.
The loss of an electron will decrease the scattering power by 17\% for a carbon atom, 14\% for nitrogen and 12\% for oxygen.
{One ionization per atom also leads to atomic displacement and further degradation of the scattered signal (Figure~\ref{fig:bragg}a)}.


$\,$
Since a focused X-ray pulse will destroy the sample,  {three-dimensional (3D)} structure
determination relies on the experiment being repeatable.
%
Rather than building up a complete X-ray diffraction data
set by rotating the crystal
and collecting a sequence of diffraction images 
it will be necessary
to scale together individual diffraction images
from many different nanocrystals,  in order to assemble a complete
3D data set~\cite{Kirian2010a}. 
A crystal with
5$\times$5$\times$5 unit cells will produce a discrete diffraction pattern~\cite{Neutze2000a},
and conventional X-ray { analysis} {techniques} may  be used for { indexing, merging and} reconstruction~\cite{Kirian2010a}. 
Furthermore, oversampling techniques for 
direct phase retrieval may also be employed for
a 3D structural determination~\cite{Miao2001a}.

We express damage-induced errors in terms of degradation of Bragg peaks, and for the entire diffraction pattern we calculate an 
{R-value} from simulated crystals exposed to X-ray pulses (Figure~\ref{fig:bragg}).  The 
{R-value} is a measure of the overall agreement between the crystallographic 
model and the experimental X-ray diffraction data (Equation~\ref{eq:rfactor}). Small molecules (such as urea) form more ordered crystals and an {R-value} below {0.05} is considered a good threshold (Cambridge Structural Database). For {macromolecules}, values up to {0.20} are acceptable (Protein Data Bank) and we use the convention  R$<0.15$ from~\cite{Neutze2000a} (Figure~\ref{fig:bragg}). Figure~\ref{fig:bragg}a shows how Bragg peaks may degrade during exposure to an X-ray pulse, and how this influences the dependence of the {R-value} with pulse parameters. The {R-value} is also dependent on crystal size, and Figures~\ref{fig:bragg}b,c compare two regimes: crystals where Auger {cascades} dominate versus crystals where damage is driven by photoelectron cascades. Trapping of photoelectrons in {crystals larger than 500 nm} leads to a steeper degradation of the signal and constrains what pulse lengths and intensities could be used for successful imaging.

When deciding which parameters of the X-ray laser pulse and sample characteristics one
should use (Figure~\ref{fig:intensity}), there is an interplay between two effects driven mainly by photoelectrons;
i) If photoelectrons escape the sample, the total number of ionizations will be significantly lowered. To reduce radiation damage {early in the exposure}, 
the sample has to be smaller than the size of the photoelectron cascade. The diffraction signal from the crystal scales with the size of the crystal as {a power law} (Equation~\ref{eq:intensity}). 
Thus reducing sample size will conversely require an increase in pulse intensity in order to {retain the signal at the same level} (Figure~\ref{fig:intensity}).
ii) If the pulse is very short, the photoelectric cascade will not have time to develop to reach a large number of ionizations. 
{At} the same time, short pulses considerably { reduce} the signal degradation due to atomic displacements and ionization (low {R-value} in
Figure~\ref{fig:bragg}). In this case the sample size is less important, and one can investigate 
{any { crystal} size at photon fluences that 
will provide enough signal. The size could however 
be constrained by coherence requirements due to the pulse length~\cite{Hau-Riege08}. At extremely short pulses, one would also need to consider the broadening of Bragg peaks~\cite{Tomov1998a}, {{\em i.e.} a bandwidth
effect}.

\begin{figure}[t]
  \begin{center}
\vspace{-1cm}
\includegraphics[width=1.0\columnwidth]{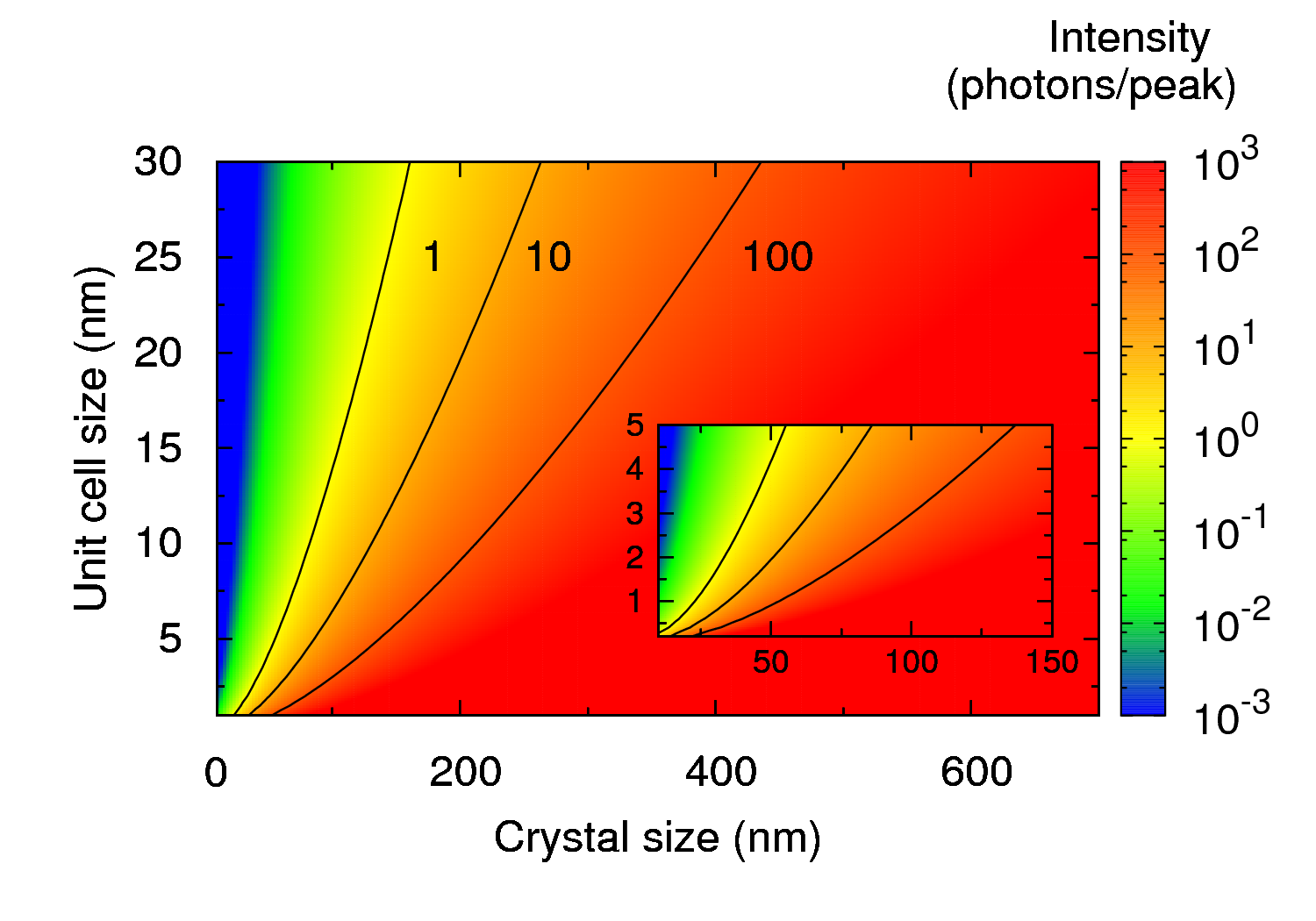}
    \caption{{\bf Photon signal as a function of unit cell and  crystal size for a perfect crystal.}  The integrated Bragg peak intensity {(Equation~\ref{eq:intensity})} is shown for the lowest resolution where a full reflection can be recorded {(Equation~\ref{eq:qmin})}. Fully integrated Bragg peaks can be used for indexing and averaging the signal from different nanocrystal orientations in the beam, and 10 detected photons are considered to give a good signal-to-noise ratio.  The X-ray pulse has an intensity of  10$^{11}$ photons focused in a focal spot of 1 $\mu$m diameter (FWHM), wavelength 1.5~{\AA}, beam divergence is 0.1 mrad and spectral bandwidth is $\Delta\lambda/\lambda = 0.1 \%$.
 The  solid lines correspond to a scattered signal of 1, 10 and 100 photons in the first fully integrated Bragg peak, when peak degradation is not taken into account.  The signal scales with the X-ray fluence, thus the  "1" line will correspond to 10 scattered photons for the case of 10$^{12}$ incident photons, and the  "100" line will correspond to 10 scattered photons for the case of 10$^{10}$ incident photons. If larger bandwidth or divergence is expected, full Bragg peaks can be recorded at lower resolutions  and consequently the detected signal will be higher (Equation~\ref{eq:intensity}). The inset shows details for crystals with small unit cells.}
    \label{fig:intensity}
  \end{center}
\end{figure}

The above  considerations stress the importance of having very short pulses as means for radiation damage control, 
to reduce both the ionization cascades and the atomic disorder. { A photon flux of 
$10^{12}$ photons per pulse} and unit area ($\mu$m$^2$) will offer the opportunity to investigate a wider range of crystal sizes and unit cells sizes. 
For lower available intensities ($10^{10}$ photons/pulse), longer pulse lengths can be accommodated and 
imaging nanocrystals of small proteins with a small unit cell, such as lysozyme, could be feasible.
{ Our calculations show that to achieve an R-value of 0.15 at a fluence 
of 10$^{11}$ photons/$\mu$m$^2$ pulses have to be shorter than 10 fs for crystals 
larger than 500 nm, where as for small crystals ($< 200$ nm) pulse lenghts can be as long as 30 fs.}

Ultrafast single-shot nanocrystallography {fills the gap} between single molecule imaging and crystallography.
It offers the opportunity to investigate  biological molecules which are too small to provide 
a good signal on their own in an X-ray laser diffraction experiment, however they could form 
nanocrystals which would be too small to investigate with conventional synchrotron radiation. 

\section*{Methods}
{\bf 1. Electron impact ionization.}
Simulations of the ionization
cascade dynamics in crystalline urea (CO(NH$_2$)$_2$)  
were performed using the spatial
electron dynamics program, \textsc{ehole}, that is a part of the
\textsc{gromacs}~\cite{Spoel2005a} Molecular Dynamics software package.
Urea
was chosen as model for a biological sample for three
reasons: it has a well
known crystalline
structure, 
it has an atomic composition of biological character, and its unit cell
is small. 
In earlier work~\cite{Caleman2009a}, the inelastic
electron cross sections for urea {have} been derived from first principles calculations.
Based on these we have calculated the number of secondary
electrons generated by an impact electron in a urea crystal. 
The inelastic cross section for electron scattering in urea is comparable in magnitude with that for water~\cite{Caleman2009a}. Thus, urea crystals are a good model for protein nanocrystals, known to contain 30\%-60\% water.
We refer to \cite{Caleman2009a,Ortiz2007a,timneanu2004a} and the supplementary
material for
further details {of} these calculations and how the model compares with experiments on diamond~\cite{Gabrysch2008a}.
Considering $\displaystyle m_i$ to be the mass of electron $i$ and $\mathbf{r}_i$ the position of electron
 $i$ with respect to the center of mass of all free electrons, 
the radius of gyration, used in Figure~\ref{fig:ionAUGvsPHOTO}, is defined as
\begin{equation}
\label{eq:rg}
R_{\mathrm{g}}(t) =\left(\frac{\sum_i \mathbf{r}_i(t)^2 m_i}{\sum_i m_i}\right)^{1/2}.
\end{equation}

\noindent{\bf 2. Electron thermalization during the pulse.}
We assume that the X-ray pulse can be described by a Gaussian centered at time $t_0=0$ and 
will consider the incoming X-ray photons to be unpolarized and have a wavelength of 1.5~\AA.
Following this pulse, several primary ionizations are treated --
the photoelectric effect resulting in an ejection of a high energy
electron ($\approx$~8 keV), accompanied by an Auger effect which provides an electron of a lower energy,
depending on atomic species. The emission for these electrons is described by normalized probability
distributions: (i) the photoelectric effect is instantaneous so the emission probability follows the same Gaussian profile as
the X-ray pulse, with the width $w$; (ii) the probability for an Auger process to be emitted
 is a convolution of a Gaussian with
the exponential decay characteristic for each individual atomic species. 
The exponential decays are taken with corresponding life times $\tau$ of 11.3~fs for carbon,
8.3~fs for nitrogen and 6.6~fs for oxygen.
The probability for photoionization in urea is determined by the
cross section of the atoms, which are well known 
For the three atomic species that can undergo an Auger process,
the contribution from the atoms
C, N, and O, is weighted according to the photoionization cross section on the
respective atoms, $\sigma_{\mathrm{C}}$,
$\sigma_{\mathrm{N}}$, $\sigma_{\mathrm{O}}$, and normalized to the total photoelectric
cross section for the urea molecule. 
The single electron ionization cascades develop mainly along the direction of the incident photon,
however we consider spherical symmetry  when these are stochastically produced. 
Thus, the entire ionization cascade following an X-ray pulse impinging on a crystal  
is given by
%
\begin{equation}
\label{eq:convolution}
\begin{split}
C(t) = & \sum_{i=\mathrm{C,N,O}} n_i\sigma_i \left\{
\int  N {\mathrm e}^{-\frac{(t'-t_0)^2}{2w^2}} C_{\mathrm{photo}}(t,t')dt'+\right. \\
 + & \left.
\int N {\mathrm e}^{-\frac{(t'-t_0)^2}{2w^2}} \frac{1}{\tau}{\mathrm e}^{-\frac{(t-t')}{\tau}}C_{\mathrm{Auger}}(t,t') dt'\right\},
\end{split}
\end{equation}
where $C_{\mathrm{photo}}(t,t')$ and $C_{\mathrm{Auger}}(t,t')$ represent the cascade 
development with time for a single
electron starting from time $t'$. These are obtained from MD simulations and are represented by the
ionization rate as a function of time (Figure~\ref{fig:ionAUGvsPHOTO}a for $t'=0$), or radii of gyration, Figures~\ref{fig:ionAUGvsPHOTO}b,c.

\noindent{\bf 3. X-ray interactions and damage quantification.}
The degradation of Bragg peaks in Figure~\ref{fig:bragg} has been calculated  from MD simulations 
on an urea crystal, using \textsc{gromacs} with {a} stochastic interaction of X-ray photons with atoms,
assuming unpolarized X-rays and homogeneous spatial distribution of the free electrons. The simulation box was 10x10x10 unit cells of urea, with periodic boundary conditions, and includes thermal motion of atoms.
The model is described in reference~\cite{Neutze2000a}, and in the supplementary material. 
The intensity of Bragg peaks at each time step is defined by integrating around each peak over 
a rectangular area centered on the Bragg peak and with sides
of length equal to 1/10 of the separation between adjacent peaks~\cite{spoel2008a}. The spectral width
$\Delta\lambda/\lambda$, beam divergence or any broadening of the Ewald sphere are not taken into account here. The degradation of the Bragg peak is expected to be smaller when integrating through 
an Ewald sphere of finite thickness.
To estimate the damage induced error we make use of the 
{R-value} (used in Figure~\ref{fig:bragg}), calculated up to a resolution $\mathbf{q}$ from
\begin{equation}
\label{eq:rfactor}
R(q) =\frac{\sum_{hkl<q} \left|  \sqrt{\langle I_{hkl} \rangle_t} -\sqrt{I^0_{hkl}} \right|}{\sum_{hkl<q} \sqrt{\langle I_{hkl} \rangle_t}},
\end{equation}
where the summation is performed over all Bragg peaks $(hkl)$ corresponding to scattering vectors less than $q$. The intensities of the Bragg peaks $I^0_{hkl}$ for the undamaged crystal are used as reference when compared with the time averaged intensities $\langle I_{hkl} \rangle_t$ of the damaged crystal exposed to a Gaussian-shaped  X-ray pulse. The latter intensities take into account the ionization dynamics and atomic displacement as a function of time during the X-ray pulse.

\noindent{\bf 4. Minimum required signal.}
In ultra-fast single-shot experiments at X-ray lasers the crystals will be exposed in random orientations, X-rays beams are expected to have small divergence ($<1$ mrad) and are highly monochromatic (spectral bandwidth $\Delta\lambda/\lambda < 0.1 \%$).
Thus, single shot diffraction patterns will contain many partially reflected Bragg peaks. Full Bragg peaks may be recorded on the detector at higher resolution, and could be used for retrieving the original orientation of the nanocrystals and for averaging the signal from similar orientation~\cite{Kirian2010a} 
(partial Bragg peaks could in principle also be used for indexing).
In our estimates for the minimum required signal for successful indexing, we consider only the signal from fully integrated Bragg peaks at the lowest resolution where these can be recorded. For given parameters that control the thickness of the Ewald sphere (beam divergence $\Delta\phi$, spectral bandwidth $\Delta\lambda/\lambda$), the lowest resolution where a full Bragg peak can be recorded ($q_{\mathrm{min}}$) is found by comparing the thickness of the Ewald sphere to the size of the Bragg peak (modeled as inversely proportional to the crystal width $A$)
\begin{align}\label{eq:qmin}
\frac 1 A \sim \frac{\Delta\lambda}{\lambda} \frac{\lambda}{2} q_{\mathrm{min}}^2   + 
\Delta\phi \, q_{\mathrm{min}} \sqrt{1-\frac{\lambda^2}{4}q_{\mathrm{min}}^2}
\,.
\end{align}

The average number of photons scattered elastically by a protein
crystal within a Bragg peak for a given resolution $q_{\mathrm{min}}$ can be approximated by the expression for the integrated reflected intensity~\cite{Kalman1979a} 
\begin{align}\label{eq:intensity}
\begin{split}
\mathrm{I_{Bragg}} (q_{\mathrm{min}})\approx
I_0(\lambda)
\frac{\lambda^2}{2\sin^2(\theta_0/2)}
\frac{1+\cos^2\theta_0}{2}\times
\\\times
\lambda^2\frac{A^3}{a^6}
r_{\mathrm{e}}^2
\sum_{\mathrm{{\mathrm{atoms}}}}f_{\mathrm{atom}}^2(\theta_0)
\,,\end{split}
\end{align}
where  both small beam divergence and polychromaticity are accounted for.  
$I_0(\lambda)$ is the spectral intensity of the incoming X-ray beam (number of incoming photons per unit wavelength and unit area), integrated over the angular density of the incident beam. The Lorentz factor $\lambda^2/(2\sin^2(\theta_0/2))$ takes into account the integration over the thickness of the Ewald sphere encompassing the Bragg peak, in a similar way as  \cite{Kalman1979a} for the stationary case (no rotation) with divergent polychromatic radiation.
The polarization factor is given by $(1+\cos^2\theta_0)/2$.
Furthermore, $r_{\mathrm{e}}$ is the classical electron radius, $\lambda$ the
wavelength, $A$ the crystal {side (cubic crystal)}, $a$ the unit cell {side (cubic unit cell)},
$f_{\mathrm{atom}}$ the atomic scattering factor, and
$\theta_0$ the polar angle between the incident pulse and the center of
the Bragg peak. 

It is assumed that the unit cell structure factor is
constant within the Bragg peak, and that adjacent Bragg peaks do not
overlap; both these approximations improve with the ratio $A/a$. The squared
structure factor of the unit cell is represented by its average value at
high scattering angles \cite{Wilson1949}. For numerical evaluation, the
unit cell was assumed to have a density of 1/30 {\AA}$^{-3}$
carbon-equivalent atoms (corresponding to a unit cell consisting of 50\%
non-structural water and protein with density
approximately 1.35 g/cm$^3$), and the scattering factor of carbon was
calculated from the Cromer-Mann parameters~\cite{Cromer1968a}.
The calculation of the number of photons per Bragg peak (Equation~\ref{eq:intensity}) presented in Figure~\ref{fig:intensity} with the assumption  of perfect crystals {with no mosaicity}. It has been shown that micron sized crystals could consist of only a 
few highly ordered domains~\cite{Fourme1995a}, thus
nanocrystals are unlikely to be organized with a mosaic spread.

The above approximations will break down when the crystal size approaches the unit cell size, as the diffracted image turns from a discrete into a continuous pattern.

\section*{Acknowledgments}
The Swedish Research Council is acknowledged for financial support as
well as the DFG Cluster of Excellence: Munich-Center for Advanced
Photonics. The authors would like to thank Magnus Bergh,
Gerard Kleywegt, Inger Andersson, Erik G. Marklund, Richard Neutze, Martin Svenda, Rosmarie Friemann, Jochen Hub 
and Karin Valeg\aa rd for their valuable input.



\end{document}